\documentclass[runningheads]{llncs}
\usepackage{amssymb}
\usepackage{latexsym}
\usepackage{graphicx}
\usepackage{times}
\usepackage{xspace}
\usepackage{amsmath}
\usepackage{enumerate}
\usepackage{oldgerm}
\usepackage[ruled,vlined]{algorithm2e}
\usepackage{multirow}
\usepackage{a4}

\newcommand{\limp}[1][]{\ensuremath{\rightarrow_{#1}}\xspace}
\renewcommand{\land}[1][]{\ensuremath{\wedge_{#1}}\xspace}
\renewcommand{\lor}[1][]{\ensuremath{\vee_{#1}}\xspace}
\newcommand{\lneg}[1][]{\ensuremath{\neg_{#1}}\xspace}
\newcommand{\when}{\mbox{$\,$:--\ }\xspace}

\newcommand{\with}[2]{\ensuremath{#1\,\mathbf{with}\,#2}\xspace}
\newcommand{\withis}{\ensuremath{\twoheadleftarrow}}

\newcommand{\insp}{\ensuremath{\,\dot{\in}\,}\xspace}

\newcommand{\dneg}[1][]{\ensuremath{\sim_{#1}\!\!}\xspace}
\DeclareMathOperator{\lub}{lub}
\DeclareMathOperator{\glb}{glb}
\newcommand{\myslash}{\mathbin{\rotatebox[origin=c]{-20}{\big/}}}
\newcommand{\mybackslash}{\mathbin{\rotatebox[origin=c]{20}{$\backslash$}}}

\newcommand{\kleene}{\ensuremath{K_3^+}\xspace}
\newcommand{\priest}{\ensuremath{P_3^+}\xspace}
\newcommand{\belnap}{\ensuremath{B_4^+}\xspace}
\newcommand{\msv}{\ensuremath{L_4^+}\xspace}

\newcommand{\defeq}{\stackrel{\mathrm{def}}{=}}
\newcommand{\defequiv}{\stackrel{\mathrm{def}}{\equiv}}

\newcommand{\tvaln}[1][]{\ensuremath{\mbox{\textswab{n}}_{#1}}\xspace}
\newcommand{\designated}[1][]{\ensuremath{\mbox{\textswab{d}}_{#1}}\xspace}
\newcommand{\tvalf}{\ensuremath{\{\,\false, \unknown, \incons, \true\}}\xspace}
\newcommand{\tvalt}{\ensuremath{\{\,\false, \unknown, \true\}}\xspace}

\newcommand{\bbd}{\ensuremath{\mathbb{D}}\xspace}
\newcommand{\bbi}{\ensuremath{\mathbb{I}}\xspace}
\newcommand{\bbl}{\ensuremath{\mathbb{L}}\xspace}
\newcommand{\bbt}{\ensuremath{\mathbb{T}}\xspace}

\newcommand{\call}{\ensuremath{\mathcal{L}}\xspace}

\newcommand{\calh}{\ensuremath{\mathcal{H}}\xspace}
\newcommand{\cali}{\ensuremath{\mathcal{I}}\xspace}
\newcommand{\calj}{\ensuremath{\mathcal{J}}\xspace}
\newcommand{\calk}{\ensuremath{\mathcal{K}}\xspace}

\newcommand{\calp}{\ensuremath{\mathcal{P}}\xspace}

\newcommand{\false}{\ensuremath{\mathbf{f}}\xspace}
\newcommand{\true}{\ensuremath{\mathbf{t}}\xspace}

\newcommand{\unknown}{\ensuremath{\mathbf{u}}\xspace}
\newcommand{\incons}{\ensuremath{\mathbf{i}}\xspace}

\newcommand{\final}[1]{#1}
\newcommand{\anonymous}[1]{}

\newcommand{\trojkat}{\footnotesize$\lhd$}
\newcommand{\done}{\hspace*{\fill}
   \mbox{\hspace*{\fill}\trojkat\break}}

\newcommand{\asp}{ASP\xspace}
\newcommand{\fsp}{\textsc{4sp}\xspace}
\newcommand{\fql}{\textsc{4ql}\xspace}
\newcommand{\dlog}{\textsc{Datalog}\xspace}

\newcommand{\saves}[1]{\ensuremath{\mbox{\em willSave}(#1)}}
\newcommand{\evacuable}[1]{\ensuremath{\mbox{\em evacuable}(#1)}}

\newcommand{\reachable}[1]{\ensuremath{\mbox{\em reachable}(#1)}}
\newcommand{\canreach}[1]{\ensuremath{\mbox{\em can\_reach}(#1)}}
\newcommand{\resc}{\mbox{\em resc}\xspace}
\newcommand{\base}{\mbox{\em base}\xspace}
\newcommand{\jack}{\mbox{\em jack}\xspace}
\newcommand{\eve}{\mbox{\em eve}\xspace}
\newcommand{\chris}{\mbox{\em chris}\xspace}
\newcommand{\interlace}[1]{\ensuremath{\mathit{interlace}(#1)}}

\newcommand{\ieeeonly}[1]{\ifieee #1\fi}
\newcommand{\aaaionly}[1]{\ifieee\mbox{}\else #1\fi}

\SetAlCapSkip{1mm}
\SetAlFnt{\small\rm}

\newenvironment{program}[1][htb]
{\renewcommand{\algorithmcfname}{Program}
	\begin{algorithm}[#1]\footnotesize
	}{\end{algorithm}%
}

\renewcommand{\tcc}[1]{{\footnotesize \ /*\ #1 \ */}}

\LinesNumbered
\DontPrintSemicolon
\newif\ifieee
\ieeefalse

\begin{document}
\frontmatter               
\pagestyle{headings}  
\addtocmark{Entailment Procedure for Answer Set Programs} 

\mainmatter  
\title{A~Paraconsistent ASP-like Language\\ with Tractable Model Generation\final{\thanks{This work has been supported by grant 2017/27/B/ST6/02018 of the National Science Centre Poland, the ELLIIT Network Organization for Information and Communication Technology, Sweden, and the Swedish Foundation for Strategic Research FSR (SymbiKBot Project).}
}}
\titlerunning{A~Paraconsistent ASP-like Language with Tractable Model Generation}  
%
\final{\author{Andrzej Sza{\l}as}}
\anonymous{\author{Anonymous}}
\final{\authorrunning{A. Sza{\l}as}}       
%
\final{\tocauthor{Andrzej Sza{\l}as (Link\"oping University) }}
\final{\institute{Institute of Informatics, University of Warsaw\\ Banacha 2, 02-097 Warsaw, Poland\\[0.5mm]
		   and\\
		Department of Computer and Information Science \\
           Link\"oping University,
           SE-581 83 Link\"oping, Sweden \\[0.5mm]
           \email{andrzej.szalas@\{liu.se, mimuw.edu.pl\}}}
}
\maketitle                   

\begin{abstract}
Answer Set Programming (\asp) is nowadays a dominant rule-based know\-ledge representation tool. Though  existing \asp variants  enjoy efficient implementations, generating an answer set remains intractable. The goal of this research is to define a~new \asp-like rule language, \fsp, with tractable model generation. The language combines ideas of \asp and a~paraconsistent rule language \fql.  Though \fsp shares the syntax of \asp and for each program all its answer sets are among \fsp models, the new language differs from \asp in its logical foundations, the intended methodology of its use and complexity of computing models. 

As we show in the paper, \fql can be seen as a paraconsistent counterpart of \asp programs stratified with respect to default negation. Although model generation of well-supported models for \fql programs is tractable, dropping stratification makes both \fql and \asp intractable. To retain tractability while allowing non-stratified  programs, in \fsp we introduce trial expressions interlacing programs with hypotheses as to the truth values of default negations. This allows us to develop a~model generation algorithm  with deterministic polynomial time complexity. 

We also show relationships among \fsp, \asp and \fql.
\end{abstract}

\section{Introduction and Motivations}
Answer Set Programming (\asp)~\cite{AlvianoFLPPT10,Baral:2003,BrewkaET11,gekakasc12a,GelKah:2014,Gelfond88thestable,Janhunen18a,Leone:2006,LinZ04,SimonsNS02} is a~knowledge representation framework based on the logic programming and nonmonotonic reasoning paradigms that uses an answer set/stable model semantics for logic programs. Generating answer sets is intractable which is both an \asp strength and a~weakness. The strength arises from concise representations of NP-complete problems and the use of efficient \asp solvers to conquer these problems.\ 
The weakness stems from potential lack of scalability: one can hardly expect efficient performance over large datasets: even generating the first answer set may require time longer than could be allocated. 

Another research line is represented by \fql~\cite{4qloxford,4qljancl,4qlamsta,4qligpl}, a~four-valued paraconsistent rule language with tractable model generation and query answering. The language allows for disambiguating inconsistencies and reacting on ignorance in a~nonmonotonic manner. For that purpose inspection operators for accessing truth values of literals have been introduced. However, tractability comes at a~price of stratification over inspection operators.  
While the \asp semantics is basically three-valued with truth values \true (true), \false (false) and \unknown (unknown)~\cite{ds2015,Przymusinski90}, \fql uses the fourth truth value, \incons, representing inconsistency. 

Paraconsistent and paracoherent versions of logic programs and \asp have been investigated in the literature~\cite{AlcantaraDP05,BlairS89,DP98,Eiter:2010,Fitting85,Fitting93}. However, to our best knowledge, no version of \asp enjoys tractable model generation. Many approaches use the  logic $B_4$~\cite{belnap} as the base formalism. However, $B_4$ may be problematic when used in the contexts we consider. Therefore, in \fql and \fsp we use the \msv logic not sharing  less intuitive features of $B_4$.\footnote{For logics used in this paper see Table~\ref{tab:logused}. The superscript `$+$' indicates that original logics are extended by introducing additional connectives.} 

In order to motivate the use of a~paraconsistent approach and the choice of  \msv rather than \belnap, consider sample rules of an imaginary rescue scenario listed as Program~\ref{prog:intro}, where \resc abbreviates ``rescuer'' and one is primarily interested in checking who is going to  be saved by the rescuer, as specified in Lines~\ref{line:p1-1}--\ref{line:p1-2} of the program.

\begin{program}[ht]
	$\saves{\resc,P}\ \ \ \ \when\, \lneg \saves{P,P}, \evacuable{P}.$\label{line:p1-1}\;
	$\lneg\saves{resc,P}\when\ \saves{P,P}.$\label{line:p1-2}\;
	$\saves{\eve,\eve}.\ \ \ \ \ \ \ \ \,  \evacuable{\eve}.$\;
	$ \lneg \saves{\jack,\jack}. \ \ \    \evacuable{\jack}.$\;
	$\lneg \saves{\resc,\resc}.\ \ \  \evacuable{\resc}.$
	\caption{Sample rules  of the rescue scenario.\label{prog:intro}}	
\end{program}

\noindent Program~\ref{prog:intro}, derived from the barber paradox, has no consistent models. Indeed, the least set of its conclusions contains, among others,

\parbox{8.3cm}{
	{\fontsize{9.5}{9}\selectfont
		\begin{align}
		&\saves{\resc,\resc}, &&\lneg\saves{\resc,\resc},\label{eq:exinc1}\\[-1mm]
		&\saves{\eve,\eve}, &&\lneg\saves{\resc,\eve},\label{eq:exinc2}\\[-1mm]
		&\lneg \saves{\jack,\jack}, &&\saves{\resc,\jack}.\label{eq:exinc3}
		\end{align}
	}
}

Despite the inconsistency in~\eqref{eq:exinc1}, conclusions in~\eqref{eq:exinc2}--\eqref{eq:exinc3} provide useful information about \eve and \jack. Of course, there may be more victims for whom conclusions are consistent. In fact, given that \evacuable{P} is consistent,  \saves{P} is consistent for all $P$ other than \resc. Importantly, inconsistent conclusions may be useful as well. First, they may indicate problematic situations calling for further attention. Second, when a~generated plan makes a~given goal inconsistent, executing the plan may be a better choice than doing nothing. E.g., if the goal is important, like helping victims, a plan with inconsistent goal may be better than having no plan. For many further arguments towards paraconsistency see~\cite{abe,akama,bertossihs,beziau,paraconsbook,shadowing,bdksz-annals,bona-hunter,carvalho,alina,tanaka,zamansky} and numerous references there.

To illustrate the questionable features of $B_4$, assume that $\saves{\chris,\chris}$ is unknown and  \saves{\resc,\resc} is inconsistent. In such a case, in $B_4$ we have:
		\begin{align}
		(\saves{\resc,\resc}\lor \saves{\chris,\chris}) \mbox{ is true};\label{eq:belnalres1}\\[-1mm]
		(\saves{\resc,\resc}\land \saves{\chris,\chris}) \mbox{ is false}.\label{eq:belnalres2}
		\end{align}

The results~\eqref{eq:belnalres1}--\eqref{eq:belnalres2} may be misleading to users sharing the classical understanding of $\lor$ and $\land$, where one expects disjunction to be true (respectively, conjunction to be false) only when at least one of~its arguments is true (respectively, false). In \msv, the disjunction in~\eqref{eq:belnalres1} is inconsistent and the conjunction in~\eqref{eq:belnalres2} is unknown. 
Consequently, we chose \msv and  \fql, adjusting the related algorithms to our needs. 

The original contributions of the paper include:
\begin{itemize}
	\item a synthesis of \asp and \fql: to design a~new language, \fsp, with tractable model generation and capturing all queries computable in deterministic polynomial time;
	\item a generalized concept of stratification: to achieve the uniformity of presentation and comparability of~\asp and \fql programs;
	\item a concept of trial expressions allowing for setting hypotheses: to accomplish tractability of \fsp model generation;
	\item Algorithm~\ref{alg:wsmfqa}: for generating well-supported \fsp models;
	\item Theorems \ref{thm:asp4ql}, \ref{thm:fqatractable}
	--\ref{thm:fqainsp}: to show relationships among \asp, \fql and \fsp as well as properties of \fsp.
\end{itemize}

The paper is structured as follows. In Section~\ref{sec:usecases} we outline the methodology behind \fsp and discuss  selected use cases of \fsp. In Sections~\ref{sec:34vl}
--\ref{sec:fql} we recall the three- and four-valued logics considered in the paper as well as the \asp and \fql languages. In Section~\ref{sec:fqa} we introduce trial expressions and the \fsp language, and present an algorithm for \fsp model generation. Section~\ref{sec:propfqa} is devoted to properties of \fsp  and its relations to \asp and \fql. Finally, Section~\ref{sec:relwork} discusses related work and concludes the paper.

\section{Methodology and Selected Use-Cases}\label{sec:usecases}

Many application examples of paraconsistent reasoning are discussed, e.g., in~\cite{abe,akama,bertossihs,beziau,carvalho,tanaka,zamansky}. \fsp can serve as a~pragmatic tool for paraconsistent reasoning in most application domains and scenarios addressed there. Below we outline the intended methodology of its use and present some further selected use cases.
\subsection{The Intended Methodology}\label{sec:intended-meth}

As we will show in Section~\ref{sec:propfqa}, every \fsp program may have an exponential number of models. However, computing each model is tractable. We therefore replace the \asp ``generate-and-test'' methodology, particularly suitable for solving NP-complete problems, by the ``generate-choose-and-use'' methodology, where one:
\begin{itemize}
	\item generates as many \fsp models as possible given particular time restrictions; 
	\item selects the best models with respect to some externally defined criteria.
\end{itemize}
Model generation may be ``blind'' if no further information is available. With additional external knowledge it may be better directed. For example, generating literals in models may be directed by suitable probability distributions when available. Given a~nonmonotonic reasoning support, one may tend to avoid {\em abnormal} literals, use defaults or results obtained from other nonmonotonic  techniques.\footnote{For a review of tractable versions of such forms of reasoning, compatible with our approach, see~\cite{4qloxford}.}

The criteria of selecting ``the best'' models are dependent on the particular goals to be achieved. E.g., one may choose models:
\begin{itemize}
	\item minimizing the number of inconsistent literals $r(.)$ for specific $r$'s;
	\item minimizing the number of unknown literals $r(.)$ for specific $r$'s;
	\item minimizing the resource consumption, cost, risks involved, etc.; 
	\item maximizing the probability of success or preferences' satisfaction;
	\item etc. 
\end{itemize}
Note that \fsp does not provide specific means for expressing such criteria. It is meant to generate models specified by programs and then to supply them for evaluation, choice and use to other systems' components. 

\begin{remark}
	Let us emphasize that in \fsp generating a model is tractable. This contrasts with \asp, where generating each model is intractable. Generating an \fsp model depends on first generating a set hypotheses (being tractable) and then using the set hypotheses to generate a model. For each set of hypotheses there is a \fsp model. Even though iterating through all sets of hypotheses is infeasible (requiring an exponential time), the intended methodology assumes generation of as many models as possible with the guarantee that the assumed (feasible) number of models will be generated. This is not the case in \asp. One may generate candidate answer sets and, assuming P$\not=$NP, finding even the first answer set may require an exponential number of iterations.\done
\end{remark}

\subsection{Selected Use-Cases}

\subsubsection{Big Data Analytics}

When big data is involved, e.g., collected from sensor networks, cyber-physical systems, IoT, health care systems, social media, smart cities, agriculture, finance, education, etc., uncertainty involving inconsistencies, noise, ambiguities and incompleteness, is inevitable~\cite{big-data}. The aim of big data analytics is to discover hidden knowledge, e.g., leading to early detection of destructive diseases or simulating risky business decisions. When rule languages are used as analytic tools, they typically use big data aggregates where inconsistencies may or have to be inherited. Enforcing consistency usually leads to loss of perhaps valuable information. For example, ``in health care systems, inconsistent information may be required to provide a~full clinical perspective where no information loss is desirable''~\cite{medicalontol}. 
Of course, when the involved facts or conclusions of rules are contradictory, the \asp consistency requirement filters out all potentially useful models.  

\subsubsection{Ontology Fusion}
\def\brain{\mbox{\em b}\xspace}\def\cns{\mbox{\em cns}\xspace}\def\bp{\mbox{\em bp}\xspace}
\def\ns{\mbox{\em ns}\xspace}
Fusing ontologies or belief bases may result in inconsistencies difficult or undesirable to recover~\cite{medicalontol,kaminskiKL15,schlobachC03}. Program~\ref{prog:ontolog} reflects the scenario discussed in~\cite{medicalontol,schlobachC03}, where two ontologies are fused and \brain, \cns, \ns, \bp stand for {\em brain}, {\em central nervous system}, {\em nervous system} and {\em body part}, respectively.

\begin{program}[ht]
	$\cns(X)\ \, \when\, \brain(X).$ \ \ \ \ \ \tcc{shared by the 1st and the 2nd ontology}\label{line:p1-1}\;
	$\bp(X)\ \;\, \when\, \brain(X).$\ \ \ \ \ \ \tcc{shared by the 1st and the 2nd ontology}\label{line:p1-2}\;
	$\ns(X)\ \ \;\when\, \cns(X).$\ \ \ \tcc{from the 1st ontology}\;
	$ \lneg \ns(X)\when\, \bp(X).$\ \ \ \ \tcc{from the 2nd ontology}\;
	$\brain(o1)$.
	\caption{Sample rules resulting from fusing ontologies.\label{prog:ontolog}}	
\end{program}
%

The set $\{\brain(o1), \cns(o1), \bp(o1), \ns(o1),\! \lneg \ns(o1)\}$ gathers conclusions of Program~\ref{prog:ontolog}. Though not an answer set, it is a \fsp model which can be used for further reasoning. While the ontology may be huge,  the inconsistency affects only literals involving $\ns(.)$. 

\subsubsection*{Some Further Use-Cases}

Due to a~limited space let us only indicate some further use cases being directly relevant to the current paper:\footnote{Please consult also references in the indicated papers.}
\begin{itemize}
	\item actions in potentially inconsistent/incomplete environments \cite{bdksz-annals};
	\item belief fusion and shadowing \cite{shadowing,shadowingmult};
	\item argumentation~\cite{alina-ijcai,alina};
	\item approximate reasoning~\cite{VMS_2009}.
\end{itemize}

\section{Many-valued Logics Used in the Paper}\label{sec:34vl}

In the rest of the paper we will focus on propositional rule languages. Of course, first-order variables are valuable as means to concisely express rule schemata.  As we consider finite domains only, our results can be lifted to the case where first-order variables are present. 

We will use three- and four-valued logics using (suitable subsets of) truth values: \false (false), \unknown (unknown), \incons (inconsistent), \true (true).\footnote{According to the standard convention we will not distinguish between truth values and constants denoting them.} To keep the presentation uniform, the syntax of all considered logics is the same: we use a set of {\em propositional variables} ({\em propositions}, for short) \calp and define formulas to be closed under unary connectives $\lneg, \dneg\;$ ({\em classical} and {\em default negation}), and binary ones $\land, \lor, \limp$ (conjunction, disjunction, implication). 

Let \call be a logic. Its set of truth values is denoted by \tvaln[\call]. We will assume that $\{\false,\true\}\subseteq\tvaln[\call]\subseteq\{\false,\unknown,\incons,\true\}$. For the purpose of defining models, we have to designate a set of truth values to act as being true~\cite{Resch,Urq}. The set of {\em designated  truth values} is denoted by \designated[\call] ($\designated[\call]\subseteq\tvaln[\call]$).

To define the semantics of \call we first need to provide the semantics of connectives applied to  truth values. Truth tables of negations and implication are shown in Table~\ref{tab:semnegimp}. 

\begin{table}\def\arraystretch{1.2}
	\caption{The semantics of negations and implication.\label{tab:semnegimp}}
	\centering
	\begin{tabular}{c||c|c}\hline
		&\lneg &\dneg \\ \hline
		\false & \true & \true \\
		\unknown & \unknown & \true \\
		\incons & \incons & \incons \\
		\true & \false & \false \\ \hline
	\end{tabular}\qquad
	\begin{tabular}{c||cccc}\hline
		\limp & \false &\unknown&\incons &\true\\ \hline
		\false & \true & \true & \true & \true \\
		\unknown & \true & \true & \true & \true \\
		\incons & \false & \false & \true & \false\\
		\true & \false & \false & \true & \true\\ \hline
	\end{tabular}
\end{table}

\begin{remark}\label{rem:negimp}
	\mbox{}
	\begin{itemize}
		\item Intuitively, the default negation $\dneg p$ as a~four-valued connective stands for ``$p$ is not true'' while the traditional \asp meaning is ``$p$~is not in the interpretation'' (when $p$ is inconsistent, both $p$ and $\lneg p$ are in the interpretation).
		\item The implication on $\{\false,\unknown,\true\}$ as well as on $\{\false,\incons,\true\}$ is the  implication of~\cite{shepherdson88}. It has been generalized to $\{\false,\unknown,\incons,\true\}$  in~\cite{msv2008b}.\done
	\end{itemize}	
\end{remark}

\begin{figure}[ht]
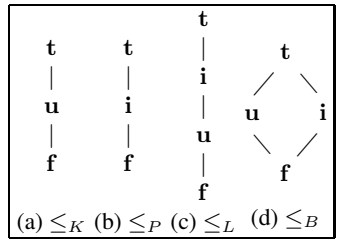
\footnotesize
	\[\begin{array}{|c|}\hline
	\begin{array}{c}
	\mbox{}\\
	\true\\
	\mid\\
	\unknown\\
	\mid\\
	\false\\
	\mbox{}\\
	\mbox{(a) $\leq_K$}
	\end{array}
	\begin{array}{c}
	\mbox{}\\
	\true\\
	\mid\\
	\incons\\
	\mid\\
	\false\\
	\mbox{}\\
	\mbox{(b) $\leq_P$}
	\end{array}
	\begin{array}{c}
	\true\\
	\mid\\
	\incons\\
	\mid\\
	\unknown\\
	\mid\\
	\false\\
	\mbox{(c) $\leq_L$}
	\end{array}
	\begin{array}{c}
	\mbox{}\\
	\true\\
	\myslash \ \ \ \ \mybackslash\\
	\unknown\ \ \ \ \ \ \ \ \ \ \incons\\
	\mybackslash \ \ \  \ \myslash   \\
	\false\\
	\mbox{}\\[-0.6em]
	\mbox{(d) $\leq_B$}
	\end{array}\\ \hline
	\end{array}
	\]
	\caption{Truth orderings used in the paper where $K$ stands for  Kleene, $P$ - Priest, $L$ - linearity, and $B$ - Belnap.\label{fig:ord}}
\end{figure}

\noindent The semantics of $\land$ and $\lor$ is standard:
{\fontsize{9.5}{9}\selectfont
	\begin{align}
	& \tau_1\land \tau_2\defeq \glb\!_{\leq}\{\tau_1,\tau_2\};\\
	& \tau_1\lor \tau_2\defeq \lub\!_{\leq}\{\tau_1,\tau_2\},
	\end{align}
}
\noindent \hspace*{-2mm} where $\tau_1,\tau_2\in\tvaln[\call]$ and $\lub, \glb$ are respectively the least upper and the greatest lower bound with respect to ordering $\leq$ chosen from Figure~\ref{fig:ord}.

To define the semantics of formulas of \call we assume a~mapping $w$ assigning truth values to propositions:
\begin{equation}\label{eq:semprop}
w:\calp_{\call}\longrightarrow\tvaln[\call].
\end{equation}
Assignments~\eqref{eq:semprop} are extended to all formulas:
{\fontsize{9.5}{9}\selectfont
	\begin{align}
	& w(\lneg A)\defeq \lneg w(A);\;\; w(\dneg A)\defeq \dneg w(A);\label{eq:extension-of-w1}\\
	& w(A\circ B)\defeq w(A)\circ w(B), \mbox{ where } \circ\in\{\land,\lor,\limp\}.\label{eq:extension-of-w2}
	\end{align}
}
Logics used in the paper are listed in Table~\ref{tab:logused}, where $K_3, P_3$ are  three-valued logics of~\cite{kleene} and~\cite{priest},  $B_4, L_4$ are four-valued logics of~\cite{belnap} and~\cite{msv2008b}, respectively. 

\begin{table}\def\arraystretch{1.2}\centering
	\caption{Logics used in the paper with underlined designated truth values.\label{tab:logused}}
	\begin{tabular}{c|c|c|c}\hline
		Logic & Extends & Truth values & Ordering\\ \hline
		\kleene & $K_3$ & \false, \unknown, \underline{\true} & $\leq_K$ in Figure \ref{fig:ord}(a)\\
		\priest & $P_3$ & \false, \underline{\incons, \true} & $\leq_P$  in Figure~\ref{fig:ord}(b)\\
		\msv & $L_4$ &  \false, \unknown, \underline{\incons, \true} & $\leq_L$  in Figure~\ref{fig:ord}(c)\\
		\belnap & $B_4$ & \false, \unknown, \underline{\incons, \true} & $\leq_B$  in Figure~\ref{fig:ord}(d)\\ \hline
\end{tabular}\end{table}  

\begin{definition}[Literals]\label{def:literal}
	Let $p\in\calp$ be a~proposition. By a~{\em classical literal} ({\em literal}, for short) we mean an expression of the form $p$ ({\em positive literal}) or $\lneg p$ ({\em negative literal}). The set of classical literals is denoted by \bbl. When $\ell\in\bbl$, $\lneg\lneg\ell$ is identified with $\ell$. By a {\em default literal} we understand an expression of the form $\dneg\ell$, where $\ell\in\bbl$. The set of default literals is denoted by \bbd.\done
\end{definition}

\begin{definition}[Interpretations, consistency]
	By an {\em interpretation} we mean a finite set $\cali\subseteq\bbl$.  An interpretation \cali is {\em consistent} if, for every $p\in\calp$, $p\not\in\cali$  or $\lneg p\not\in\cali$.
	\done
\end{definition}

In what follows, the considered set of propositions, \calp, will always be finite. In such cases there is a one-to-one mapping between assignments~\eqref{eq:semprop} and interpretations allowing us to freely switch between them. Namely, given an interpretation~\cali, the corresponding assignment $w^\cali$ is: 
\[\fontsize{9.5}{9}\selectfont
w^{\cali}(p)\defeq\left\{ 
\begin{array}{ll}
\true & \mbox{when } p\in\cali \mbox{ and } \lneg p\not\in\cali;\\
\incons & \mbox{when } p\in\cali \mbox{ and } \lneg p\in\cali;\\
\unknown & \mbox{when } p\not\in\cali \mbox{ and } \lneg p\not\in\cali;\\
\false & \mbox{when } p\not\in\cali \mbox{ and } \lneg p\in\cali.
\end{array}
\right.
\]
Of course, $w^\cali$ can be extended to all formulas using Equations~\eqref{eq:extension-of-w1} and~\eqref{eq:extension-of-w2}. To simplify notation we will write $\cali(A)$ to stand for $w^{\cali}(A)$.
Conversely, given $w$, the corresponding interpretation $\cali^w$ is defined by:\\[-1.em]
\[\fontsize{9.5}{9}\selectfont
\cali^w\!\defeq\!\{p\mid w(p)\!=\!\true\}\cup\{\lneg p\mid w(p)\!=\!\false\}\cup\{p, \lneg p\mid w(p)\!=\!\incons\}.
\]

\begin{definition}[Models]
	Given a~logic \call with the set of designated values $\designated[\call]\!\subseteq\tvaln[\call]$, we say that an interpretation \cali\ {\em is a~model} of a~formula $A$, $\cali\models_{\call} A$, when  $\cali(A)\in \designated[\call]$. \done
\end{definition}

\section{Answer Set Programming}\label{sec:asp}

We will focus on {\em normal} \asp programs.\footnote{The results can be extended to disjunctive programs where disjunctions correspond to choice rules but due to the space limit, we do not consider this extension here.}  

\begin{definition}[Syntax of normal \asp programs]\label{def:asp}\mbox{}\\
	Let   $\ell_1,\ldots, \ell_k, \ell_{k+1}, \ldots\ell_m\in\bbl\cup\tvalt$  and $\ell\!\in\!\bbl$. A~{\em normal \asp rule} ({\em \asp rule}, for short) is an expression of the form:
	\begin{equation}\label{eq:normalrule}
	\ell\when\ \ell_1,\ldots, \ell_k, \dneg\ell_{k+1}, \ldots,\dneg\ell_m.
	\end{equation}
	It is further assumed that $0\leq k\leq m$ and when $k+1\leq m$ then $k>0$.\footnote{Default literals have to be ``guarded'' by a classical literal.} Literal $\ell$ is called the {\em conclusion} ({\em head}) and the part after `\!\when$\!\!\!$' is called the {\em premises} ({\em body}) of rule~\eqref{eq:normalrule}.
	A rule with the empty premises is called a~{\em fact} and is written as `$\ell.$'. A~rule without default literals is called {\em pure}.
	
	A~{\em normal \asp  program} ({\em \asp program}, for short) is a finite set of normal rules. A~program is {\em pure} if it contains pure rules only.
	\done
\end{definition}

\begin{remark}
	Note that we require a nonempty conclusion, so \asp constraints are excluded. With \asp-like constraints, tractability could be lost. Indeed, iterating through the set of hypotheses could take superpolynomial time when models could be rejected by constraints. 
	\done
\end{remark}

Note also that the empty conjunction, thus the empty body, is assumed to be \true.

The basic rule-based reasoning principle of \asp is:
\begin{equation}\label{eq:princ1}
\parbox{7.2cm}{-- if premises of rule~\eqref{eq:normalrule} evaluate to \true,\\ 
	\mbox{}\hspace*{3.mm}add $\ell$ to the~set of conclusions.}
\end{equation}

The semantics of \asp programs is given by answer sets.

\begin{definition}[Models of \asp programs; Answer Sets]\label{def:answset}
	By~a~{\em model} of an \asp program $\Pi$ we mean a~consistent interpretation \cali satisfying all rules of $\Pi$ understood as implications:
	\[\cali\models_{\kleene}(\ell_1\land\ldots\land \ell_k\land \dneg\ell_{k+1}\land \ldots\land\dneg\ell_m)\limp\ell.
	\]
	If $\Pi$ is pure then an {\em answer set of} $\Pi$ is the least (with respect to $\subseteq$) model of $\Pi$, if exists.
	If $\Pi$ contains \dneg\, then \cali is an~{\em answer set of $\Pi$} iff $\cali$ is the least model of $\Pi^\cali$, where $\Pi^\cali$ is obtained from $\Pi$ by substituting each default literal $\dneg\ell$ occurring in $\Pi$ by its truth value $\cali(\dneg\ell)$.   \done
\end{definition}  

Let us now provide a~simple (naive) algorithm for generating an answer set for an~\asp program. We first need Algorithm~\ref{alg:pure} generating minimal interpretations for pure programs. To keep presentation simple, it is based on the naive bottom up evaluation - see~\cite{AHV}. Each literal of the form $\lneg p$ is treated as a fresh propositional variable, say $p'$, so pure \asp programs can be seen as \dlog programs to which the original naive bottom up evaluation  applies. Generating an answer set can be done nondeterministically, as in Algorithm~\ref{alg:ansset}.

\begin{algorithm}[ht]
	\caption{\textbf{function} generateLeast($\Pi$); \label{alg:pure}}
	\tcc{returns the least model of a pure \asp~program $\Pi$}
	\BlankLine
	\textbf{set} $\cali=\emptyset$;\;
	\While{there is a~rule $`\ell\when \beta.\mbox{'} \in \Pi$ such that\\
		\hspace*{8mm} $\cali(\beta)=\true$ and  $\ell\not\in\cali$}
	{
		\textbf{set} $\cali = \cali \cup \{\ell\}$;\;
	}
	\textbf{return} \cali.
\end{algorithm}

\begin{algorithm}[ht]
	\caption{\textbf{function} generateAnswerSet($\Pi$); \label{alg:ansset}}
	\tcc{returns an answer set of an \asp program $\Pi$ if exists}
	\BlankLine
	\textbf{set} $\cali=$  a~nondeterministically generated consistent interpretation;\;
	\textbf{set} \calj = generateLeast$(\Pi^\cali)$;
	\tcc{$\Pi^\cali$ is defined in Def.~\ref{def:answset}}\;
	\If{\cali=\calj}{\textbf{return} \cali.}
\end{algorithm}

The following theorem is well known (see, e.g.,~\cite{Baral:2003,BrewkaET11,marek-sub}).

\begin{theorem}\label{thm:npasp}
	Given an \asp program $\Pi$,  generating an answer set for $\Pi$ is an NP-complete problem with respect to the number of propositions in $\Pi$.\done
\end{theorem}

\section{The \fql Language}\label{sec:fql}

The \fql language has been introduced in~\cite{4qloxford,4qljancl}. In this paper we shall use its extended version of~\cite{4qlamsta}.\footnote{\ieeeonly{For an~open source interpreter of \fql and its doxastic extensions (e.g.,~\cite{bdksz13,shadowing}) -- see \texttt{4ql.org}.}\aaaionly{An~open source interpreter of \fql and its doxastic extensions (see~\cite{shadowing} and references there) is available via \texttt{4ql.org}.}} \fql allows for paraconsistent reasoning, using the \msv logic. Rather than default negation \dneg, inspection operators are used as defined below.

\begin{definition}[Inspection operators]
	By an {\em inspection operator} we understand any expression of the form $\ell\insp T$ , where $\ell\in\bbl$ and $T\subseteq\tvalf$. The meaning of inspection operators depends on the actual interpretation \cali:
	\begin{equation}\label{eq:inspec2val}
	\cali(\ell\insp T)\defeq\left\{
	\begin{array}{ll}
	\true & \mbox{when $\cali(\ell)\in T$};\\
	\false & \mbox{otherwise}. 
	\end{array}
	\right.
	\end{equation}
	The set of  inspection operators is denoted by \bbi.	
	\done
\end{definition}


When truth values are restricted to three values of \kleene, default negation of \asp can be defined by:\\[-0.9em]
\begin{equation}\label{eq:dnegtoinsp}
\dneg\ell\defequiv\big(\ell\insp\{\,\false,\unknown\}\big).
\end{equation} 

The original version of~\fql uses modules but, for the sake of uniformity, we  skip them here. In order to compare \fql, \asp, and \fsp as well as to achieve the full power of \fql without using modules, let us introduce a~general form of stratification with respect to a set of arbitrary expressions $\mathbb{E}$, e.g., consisting of default literals or expressions involving inspection operators.

\begin{definition}[Stratification]\label{def:strat}
	Given a finite set $S$ of \fql (or \asp) rules and a set of expressions $\mathbb{E}$, we say that $S$ is {\em stratifiable with respect to $\mathbb{E}$} when  $S\!=\!S_1\cup\ldots\cup S_r$ such that for $1\!\leq\! i\not= j\!\leq\! r$, $S_i\cap S_j=\emptyset$~and:\\[-1.3em]
	\begin{itemize}
		\item for every conclusion $\ell$ of a rule in $S$, there is $1\!\leq\! i\!\leq\! r$ such that all rules with conclusions  $\ell, \lneg\ell$ are in $S_i$ ($\ell$ is {\em fully defined} in $S_i$);
		\item whenever an expression $e\not\in \mathbb{E}$ appears in premises of a~rule in $S_i$, for $1\leq i\leq r$, classical literals appearing in $e$ are fully defined in $S_j$ for some $1\leq j\leq i$;
		\item whenever an expression $e\in \mathbb{E}$ appears in premises of a~rule in $S_i$ for $1\leq i\leq r$,  classical literals occurring in $e$ are fully defined in $S_j$ for some $1\leq j< i$.\done 
	\end{itemize}  
\end{definition}

\begin{remark}\label{rem:strattract}
	Stratification of Definition~\ref{def:strat} generalizes stratification used in \dlog$^{\!\!\lneg}$~\cite{AHV}. To verify whether a set of \fql rules is stratifiable, one can easily adapt the  deterministic polynomial time algorithm checking stratification for \dlog$^{\!\!\lneg}$ programs. Finding a~stratification, if exists, is also tractable.
	\done
\end{remark}

\begin{definition}[Syntax of \fql programs]\label{def:fql}\mbox{}\\
	Let $\ell\in\bbl$, $\ell_{11},\ldots,\ell_{1k_1}$,$\ldots,$ $\ell_{m1},\ldots,\ell_{mk_m}\!\!\in$ $\bbl\cup\bbi\cup\tvalf$.    	
	A~{\em \fql rule}  is an expression of the following form, where semicolon `;' stands for disjunction, with conjunction `,' binding stronger:\\[-0.8em]
	\begin{equation}\label{eq:4qlrule}
	\ell  \when\ \ell_{11},\ldots, \ell_{1k_1};
	\ldots;
	\ell_{m1}, \ldots,\ell_{mk_m}.
	\end{equation}
	A~rule without occurrences of inspection operators is called {\em pure}.
	A~\fql\ {\em program} is a~finite set of \fql rules stratifiable with respect to~\bbi. A program is {\em pure} if it contains pure rules only.\done
\end{definition} 
Rule~\eqref{eq:4qlrule} is interpreted as the following implication of \msv:\\[-0.8em]
\begin{equation}\label{eq:4qlruleasimp}
\big((\ell_{11}\land\ldots\land \ell_{1k_1})\lor
\ldots\lor
(\ell_{m1}\land \ldots\land\ell_{mk_m})\big)\limp \ell.
\end{equation}
The reasoning principles of \fql are \eqref{eq:princ1} together with:
\begin{equation}\label{eq:princ2}
\parbox{7.cm}{-- if premises of rule~\eqref{eq:4qlrule} evaluate to \incons,\\ \hspace*{2mm} add $\ell,\lneg\ell$ to the set of conclusions.}
\end{equation}

\begin{remark}\label{rem:disj}
	Though Principles~\eqref{eq:princ1} and \eqref{eq:princ2} are natural, they appear problematic when disjunction is concerned. As an~example, consider Program~\ref{prog:disj}. 
	\begin{program}[ht]
		$\reachable{base,P}\when\ \canreach{base,P,ground}$.\label{line:reachable1}\;
		$\reachable{base,P}\when\ \canreach{base,P,air}.$\label{line:reachable2}
		\caption{Program illustrating a disjunction issue.\label{prog:disj}}
	\end{program}
	
	When one of premises of rules in Lines~\ref{line:reachable1}--\ref{line:reachable2} of Program~\ref{prog:disj} evaluates to \true with the other one being \incons, according to Principles~\eqref{eq:princ1}, \eqref{eq:princ2}, $\mathit{reachable(base,P)}$ becomes \incons. On the other hand,  the disjunction should intuitively be true, so one would rather expect a~rule like:
	\begin{equation}\label{eq:disj}
	\begin{array}{ll}\small
	\!\!\!\!\!\!\mathit{reachable(base,\!P)}\when&\!\!\!\!\!\! \mathit{can\_reach(base,\!P\!,\!ground)}\,\lor\\
	&\!\!\!\!\!\! \mathit{can\_reach(base,P,air)}.
	\end{array}
	\end{equation}
	In such circumstances, using rule~\eqref{eq:disj} one indeed concludes that $\reachable{\base,P}$ is \true in each logic listed in Table~\ref{tab:logused}.
	
	Accordingly, disjunction is explicit in rules. Due to its nonmonotonic behavior, it requires a~nonstandard computation engine.\footnote{A~\dlog-like evaluation applied to reducts in \asp  is far from being sufficient here.} For details see~\cite{4qloxford,4qlamsta} and further parts of the current paper.\done
\end{remark}

The semantics of \fql is defined by well-supported models, where {\em well-supported\-ness} guarantees that all conclusions are derived using reasoning starting from facts~\cite{4qloxford,4qljancl,4qlamsta}. For a~definition of well supported models see~\cite{4qljancl}. To put well-supportedness  into perspective, below we provide a~new, equivalent definition generalizing the concept of loops~\cite{LinZ04}.

\begin{definition}[Dependency graph]
	Let $\Pi$ be a~pure \fql program. By a {\em dependency graph}	of $\Pi$ we understand a directed graph with vertices labeled by classical literals occurring in $\Pi$. There is an arc from $\ell$ to $\ell'$ if there is a~rule in $\Pi$ whose head is $\ell$ and $\ell'$ appears in the rule's body.\done
\end{definition}

\begin{definition}[Loop]
	A non-empty subset $L$ of literals occurring in a pure \fql program $\Pi$ is called a {\em loop} of $\Pi$ if for any $\ell,\ell'\in L$, there is a path of non-zero length from $\ell$ to $\ell'$ in the dependency graph of $\Pi$, such that all the vertices in the path are in $L$. 
	\done
\end{definition}
By $R^-(L,\Pi)$ we understand the set of rules:\\[-0.8em]
\begin{equation}
\begin{array}{ll}
\big\{&\!\!\!\!\!\!\mbox{`}\ell\when B_1;\ldots;B_m.\mbox{'}\in\Pi \mid\ \ell\in L\mbox{ and }\mbox{there is }\\
& 1\leq i\leq m
\mbox{ such that for all } \ell' \mbox{ in } B_i, \ell'\not\in L\big\}.
\end{array}
\end{equation}

\begin{definition}[Well-supported model]
	An interpretation \cali is a {\em well-supported model} of a pure \fql program $\Pi$ iff \cali is the least (with respect to $\subseteq$) model of $\Pi$,\footnote{In the four-valued case, minimality substitutes the completion of~\cite{Clark} used together with loop formulas to characterize answer sets for \asp programs~\cite{LinZ04}.} 
	and for every loop $L$ of $\Pi$, if there is $\ell\in L$ such that $\cali(\ell)\in\{\incons,\true\}$ then:
	\begin{itemize}
		\item $\cali(\ell)=\true$ iff there is a~rule `$\ell\when B.$\mbox{'} in $R^-(L,\Pi)$ such that $\cali(B)=\true$ and there are no rules `$\ell\when C.\mbox{'}, \mbox{`}\lneg \ell\when D.\mbox{'}$ in $R^-(L,\Pi)$ such that $\cali(C)=\incons$ or
		$\cali(D)\in\{\incons,\true\}$;
		\item $\cali(\ell)=\incons$ iff there is a~rule `$\ell\when B.$\mbox{'} or `$\lneg \ell\when B.$\mbox{'} in $R^-(L,\Pi)$ such that $\cali(B)=\incons$ or 
		there are rules `$\ell\when C.\mbox{'}$, $\mbox{`}\lneg \ell\when D.\mbox{'}$ in $R^-(L,\Pi)$ with $\cali(C)\!=\!\true$ and  $\cali(D)\!=\!\true$.\done
	\end{itemize} 
\end{definition}

Algorithm~\ref{alg:wsminformal} presents a high-level pseudocode for computing well-supported models for pure \fql programs. It is further formalized as Algorithm~\ref{alg:correct} (implementing Line~\ref{line:correct} of Algorithm~\ref{alg:wsminformal}), and Algorithm~\ref{alg:wsmfql} (computing the well-supported model). Note that all conclusions inferred by Algorithm~\ref{alg:wsminformal} are supported by facts and no conclusion is obtained using a~proposition defeated later.

\begin{algorithm}
	\caption{A pseudocode for computing the well supported model for a given pure \fql program.\label{alg:wsminformal}}
	\Repeat{no further retractions are needed.}{ 
		\textbf{generate} the least set of conclusions;\;
		\textbf{retract} conclusions based on defeated premises,\break i.e., premises at some point being true but later becoming inconsistent;\;
		\textbf{correct} (minimally) the obtained set of literals to make it a~model (to satisfy rules with inconsistent premises and not inconsistent conclusions) \label{line:correct}
	}
	
\end{algorithm}

\begin{remark}\label{rem:inspelim}
	To generalize Algorithm~\ref{alg:wsmfql} to non-pure programs, one can find its stratification (without losing tractability -- see Remark~\ref{rem:strattract}) and eliminate inspection operators stratum by stratum, starting from the lowest one.\footnote{Here we understand stratification in the sense of Definition~\ref{def:strat} with $\mathbb{E}=\bbi$.} Let $S_i$ be the lowest stratum where inspection operators occurs. Their truth values are determined in strata lower than $S_i$. Substituting all inspection operators occurring in $S_i$ by the determined truth values makes $S_1\cup\ldots\cup S_i$ a~pure program for which Algorithm~\ref{alg:wsmfql} applies. This procedure is to be iterated until all strata have been considered. \done
\end{remark}

\begin{algorithm}[ht]
	\caption{\textbf{function} findCorrection($\Pi$, \cali); \label{alg:correct}}
	\tcc{returns the correction of \cali with respect to a pure \fql program $\Pi$}
	\BlankLine
	\textbf{set} $\calj=\emptyset$;\;
	\textbf{set} $\calk=\cali$;\;
	\While{there is a~rule $`\ell\when \beta.\mbox{'} \in \Pi$ such that $\calk(\beta)=\incons$ and  $\calk(\ell)\not=\incons$}{
		\textbf{set} $\calj = \calj\cup \{\ell,\lneg\ell\}$;\;
		\textbf{set} $\calk = \calk\cup \{\ell,\lneg\ell\}$;\;
	}
	\textbf{return} $\calj$.
\end{algorithm}

\begin{algorithm}[ht]
	\caption{\textbf{function} generateWsm4ql($\Pi$)\label{alg:wsmfql}}
	\tcc{returns the  well-supported model \cali for a \fql program $\Pi$}
	\BlankLine
	\SetKwBlock{Begin}{}{end}%
	\textbf{set} $Inc\!=\!\emptyset$;\tcc{$\!Inc$  is the set of inconsistent literals detected so far\!}\;
	\Repeat{$\calj=\emptyset$;}{
		\textbf{set} $PTrue$ = generateLeast$(\Pi\!\setminus\!\{\mbox{`}\ell\when\beta.\mbox{'}\mid \ell\!\in\! Inc\})$;
		\break\tcc{$PTrue$  is the set of potentially true literals}\;
		\textbf{set} $\cali=Inc\cup \{\ell\mid \ell\in PTrue\}$;\label{line:five}\;
		\textbf{set} \calj = findCorrection$(\Pi,\cali)$;\;
		\textbf{set} $Inc = \{p,\lneg p\mid \cali(p)=\incons\}\cup\calj$;\label{line:x}
	}
	\textbf{return} \cali.
\end{algorithm}

The following theorem is proved in~\cite{4qljancl,4qlamsta}.

\begin{theorem}\label{thm:4ql}\mbox{}
	\begin{itemize}
		\item For every \fql program $\Pi$ there is exactly one well-supported model and it can be generated in deterministic polynomial time with respect to the number of propositions in $\Pi$.
		\item \fql captures deterministic polynomial time over linearly ordered domains. That is, every query computable in deterministic polynomial time can be expressed in \fql whenever a~linear ordering over the domain is available in the \fql vocabulary.\done
	\end{itemize}
	
\end{theorem}

When stratification is not required, like in the case of \asp, generating well supported models for \fql programs becomes an NP-complete problem. 

The following theorem shows a close correspondence between (stratified) \asp and \fql programs.

\begin{theorem}\label{thm:asp4ql}
	Let $\Pi$ be an \asp program stratifiable with respect to the set of default literals \bbd and $\Pi'$ be a~\fql program obtained from $\Pi$ by substituting default negations by inspection operators as in~\eqref{eq:dnegtoinsp}. Then the well supported model \cali of $\Pi'$ is the  answer set of $\Pi$ iff \cali is consistent. If \cali is inconsistent then $\Pi$ has no answer set.\done
\end{theorem}

\section{The \fsp Language}\label{sec:fqa}


The syntax of \fsp programs is very similar to \asp programs. We extend the language by allowing disjunctions and the truth constant \incons to appear in rules' bodies. 

\begin{definition}[Syntax of \fsp programs]\label{def:fqa}\mbox{}\\
	Let $\ell\!\in\!\bbl$ and $\ell_{11}\!,\ldots,\!\ell_{1k_1}\!,\ldots,$ $\ell_{m1},\!\ldots,\!\ell_{mk_m}\!\!\in$ $\bbl\cup\bbd\cup\tvalf$.    	
	A~{\em \fsp rule}  is an expression of the following form, where semicolon `;' stands for disjunction, with conjunction `,' binding stronger:
	\begin{equation}\label{eq:4sprule}
	\ell  \when\ \ell_{11},\ldots, \ell_{1k_1};
	\ldots;
	\ell_{m1}, \ldots,\ell_{mk_m}.
	\end{equation}
	A \fsp rule is {\em pure} if it does not involve  default negation.
	A~\fsp\ {\em program} is a~finite set of \fsp rules. A~\fsp program is {\em pure} if it contains pure rules only.\done
\end{definition}

The key step towards tractable model generation is to use the four valued default negation, as defined in Table~\ref{tab:semnegimp}. To illustrate the idea, let us start with supportedness losing in the non-stratified case.

\begin{example}\label{ex:suploosing}
	When \fql programs may be non-stratified, supportedness may be lost. To illustrate the issue, consider non-stratifiable \fql rules in Program~\ref{prog:unk}.\footnote{Program~\ref{prog:unk} is a set of rules being neither a \fql nor a \fsp program.} 
	The program has no well-supported models: the rule in Line~\ref{line:unk1} makes $p$ true so the second rule makes $q$ true, too. That way $p$ loses its support so cannot be derived. Consequently, $q$ cannot be derived, so its value becomes \unknown and so on.
	
	\begin{program}[ht]
		$p\when\ q \insp \{\unknown,\false\}.$\label{line:unk1}\quad \tcc{p\when \dneg q.\ -- when restricted to \asp} \;
		$q\when\ p.$\label{line:unk2}
		\caption{Supportedness losing.\label{prog:unk}}
	\end{program}
	
	When Line~\ref{line:unk1} is replaced by `$p\when\!\! \dneg q.$', with the four-valued \dneg\,, after applying the rules it would be natural to consider $\dneg q$ inconsistent. Now,  Principle~\eqref{eq:princ2} together with the rule in Line~\ref{line:unk1} make $p$ inconsistent and the rule in Line~\ref{line:unk2} makes $q$ inconsistent. So the well-supported model would become $\{p,\lneg p, q, \lneg q\}$.
	\done
\end{example}

In \fsp, rather than using two-valued inspection operators, we will use the four-valued default negation \dneg\,. However, we encounter the next issue, illustrated in Example~\ref{ex:issue}.

\begin{example}\label{ex:issue}
	Consider Program~\ref{prog:issue}. It consists of two non-stratifiable rules.
	\begin{program}[h]
		$p\when\ \dneg q.$\label{line:issue1}\;
		$q\when\ \dneg p.$\label{line:issue2}
		\caption{A further non-stratifiability effect.\label{prog:issue}}
	\end{program}
	\noindent Both $\{p\}$ and $\{q\}$ are its well-supported models. Their generation depends on the order of rules: when the  rule in Line~\ref{line:issue1} is applied first,  the result is $\{p\}$, being  $\{q\}$ otherwise.    	
	\done
\end{example}  

Of course, the semantics should not depend on the order of rules' application. We therefore consider programs in the context of~{\em trial expressions} allowing one to select truth values of default negations. 

\begin{definition}[Trial expression]
	By a~{\em trial expression} we understand an expression of the form:\\[-0.8em]
	\begin{equation}\label{eq:trialexp}
	\with{\Pi}{\calh},
	\end{equation}  
	where $\Pi$ is a~program with \asp syntax in the sense of Definition~\ref{def:asp} and $\calh$, called a~{\em set of hypotheses for $\Pi$}, is a~finite set of expressions of the form $\dneg\ell\withis\true$ or $\dneg\ell\withis\false$ such that for every and only literal $\ell'$ occurring in $\Pi$ within an expression $\dneg\ell'$, the literal $\ell'$ occurs (as a subexpression) in $\calh$.\done  
\end{definition}
The set of trial expressions is denoted by \bbt.

Intuitively, \with{\Pi}{\calh} amounts to assuming the truth values of default negations in $\calh$ and verifying whether the hypotheses have been consistent with the results of $\Pi$. For a~particular default negation occurring in $\Pi$,  
\begin{itemize}
	\item first assume that the truth value $\dneg\ell$ is \true (respectively, \false);
	
	\item if, during generating an answer set for $\Pi$, the value of $\dneg\ell$ appears not to be \true (respectively, not to be \false), correct the truth value of $\dneg\ell$ assigning it the value \incons. 
\end{itemize} 
Note that we do not allow expressions of the form $\dneg\ell\withis\incons$. First, the role of trial expressions is to try to ``guess'' consistent solutions with inconsistency being an undesirable effect. Second,  
$\dneg\ell\withis\incons$  can be expressed by $\{\dneg\ell\withis\false, \dneg\lneg\ell\withis\false\}$. This follows from the fact that $\dneg\ell$ is \false only when $\ell$ is \true.

\begin{example}
	Let $\Pi$ denote  Program~\ref{prog:issue}. Then:
	\begin{equation}\label{eq:exampletrial}
	\with{\Pi}{\{\dneg q\withis\true,\dneg p\withis\false\}}
	\end{equation}
	is intended to mean that the value of $\dneg q$ is hypothesized to be true and the value of $\dneg p$ is hypothesized to be false. Hence,~\eqref{eq:exampletrial} results in $\{p\}$, confirming the hypotheses. 
	
	On the other hand, consider:
	\begin{equation}\label{eq:exampletrialtwo}
	\with{\Pi}{\{\dneg p\withis\true,\dneg q\withis\true\}}.
	\end{equation}
	The rules of Program~\ref{prog:issue} generate $\{p,q\}$ as conclusions, violating both hypotheses expressed in~\eqref{eq:exampletrialtwo}. Therefore, both $\dneg p$ and $\dneg q$ become \incons and the answer set for~\eqref{eq:exampletrialtwo} becomes $\{p,\lneg p, q, \lneg q\}$.
	\done
\end{example}

\begin{remark}\label{rem:sideeffects}
	Note that hypotheses of~\eqref{eq:trialexp} may mutually affect knowledge bases represented by programs. For example, $\lneg p$ is inconsistent with the hypothesis that $\dneg p\withis\false$. If $\lneg p$ is a~generated conclusion, $\dneg p\withis\false$ together with $\lneg p$ makes $\dneg p$ inconsistent, what only happens when $p$ is \incons. This calls for placing $p$ in the generated model, too. Such side-effects have to be reflected in the model generation algorithm.\done
\end{remark}

In importing truth values from an interpretation \cali to the set of hypotheses $\calh$ we deal with the situation when the imported truth values are not \unknown. Therefore, we are on the grounds of  \priest and can use the following rules:\footnote{The rule~\eqref{eq:interlacefirst} applies to $\ell$ being either positive or negative. }
\begin{equation}
\mbox{if } \ell\in\cali\mbox{ then add } \dneg\ell\withis\false\mbox{ to } \calh.\label{eq:interlacefirst}
\end{equation}
Of course, when $\ell$ is \incons in \cali, $\ell,\lneg\ell\in\cali$ so the rule~\eqref{eq:interlacefirst}, applied twice, results in adding to \calh both $\dneg\ell\withis\false$ and $\dneg\lneg \ell\withis\false$. 

Conversely, 
\begin{align}
& \mbox{if } \{\dneg\ell\withis\false, \dneg\ell\withis\true\}\subseteq \calh \mbox{ then add } \ell, \lneg\ell \mbox{ to }\cali;\\
& \mbox{else if } \dneg\ell\withis\false\in\calh \mbox{ then add } \ell\mbox{ to } \cali.\label{eq:interlacelast}
\end{align}

\begin{definition}[Semantics of \fsp]
	Given an \fsp program $\Pi$, by a~well-supported model of $\Pi$ we mean any well-supported model for `$\with{\Pi}{\calh}$' where $\calh$ is an~arbitrary set of hypotheses for $\Pi$.	\done
\end{definition}

For computing well-supported models of \fsp programs we still need the following definition.

\begin{definition}[Interlace]\label{def:interlaced}
	Let  \cali be an interpretation and $\calh$ be a set of hypotheses for a~\fsp program. By an {\em interlace} of \cali and $\calh$, denoted by $\interlace{\cali,\calh}$, we mean a~minimal interpretation \calj such that $\cali\subseteq\calj$ and \calj is closed under application of rules~\eqref{eq:interlacefirst}--\eqref{eq:interlacelast}.\done
\end{definition}

Algorithm~\ref{alg:wsmfqa} computes the well-supported model for a~\fsp program $\Pi$ when a set of hypotheses \calh is given. The function `generateWsmAux' is obtained from Algorithm~\ref{alg:wsmfql} by adding \calh  as a~parameter and replacing Line~\ref{line:five} by:\\[-0.5em]
\begin{equation}\label{eq:interlacein}
\mbox{\scriptsize\bf 5'\ \ }\mathbf{set}\ \cali=\interlace{Inc\cup \{\ell\mid \ell\in PTrue\},\calh}.
\end{equation}


\begin{algorithm}[h]
	\caption{\textbf{function} generateWsm4sp($\Pi,\calh$)\label{alg:wsmfqa}}
	\tcc{\footnotesize returns the well-supported model \cali for \fontsize{9}{8} `\with{\Pi}{\calh}'}
	
	
	\textbf{return} generateWsmAux($\Pi,\calh$).
\end{algorithm}

\begin{remark}
	Though Algorithm~\ref{alg:wsmfqa} may look  simple, it actually substantially differs from the algorithms for generating well-supported models for \fql programs due to the use of  \interlace{} in~\eqref{eq:interlacein} and related concepts.\done
\end{remark}

\section{Properties of \fsp}\label{sec:propfqa}

Let us now focus on the most important properties of \fsp.

\begin{theorem}\label{thm:fqatractable}\mbox{}
	\begin{itemize}
		\item For every \fsp program  $\Pi$ and every set of hypotheses $\calh$ for $\Pi$  provided as input to Algorithm~\ref{alg:wsmfqa}, the unique well-supported model is computed in deterministic polynomial time in the number of propositions in $\Pi$ and the size of \calh.
		\item \fsp captures deterministic polynomial time over linearly ordered domains.
		\done
	\end{itemize}
\end{theorem}

To generate all well-supported models for a given \fsp program $\Pi$ it suffices to iterate Algorithm~\ref{alg:wsmfqa} with sets \calh reflecting different choices of truth values \true, \false assigned to default literals in $\Pi$. 

\begin{theorem}\label{thm:fqanumber} For every \fsp program there is at most an exponential number of well-suppor\-ted models with respect to the number of default literals occurring in $\Pi$. \done	
\end{theorem}

In the light of Theorem~\ref{thm:fqanumber}, the intended use of Algorithm~\ref{alg:wsmfqa} instantiates the methodology outlined in Section~\ref{sec:intended-meth}, where model generation depends on assigning truth values to hypotheses. 

\begin{theorem}\label{thm:fqacontainasp}
	For every \asp program $\Pi$, the set of all well-supported models of $\Pi$ (being a \fsp program) contains all answer sets of $\Pi$.	\done
\end{theorem}

When inspection operators are also allowed in premises of \fsp rules, we have the following theorem.

\begin{theorem}\label{thm:fqainsp}
	When inspection operators are allowed,
	\begin{itemize}
		\item model generation is tractable for \fsp programs stratifiable with respect to \bbi;
		\item model generation is NP-complete for \fsp programs without the stratifiability requirement. \done
	\end{itemize}  
\end{theorem}

\section{Related Work and Conclusions} \label{sec:relwork}

The paper combines two threads: \asp\cite{AlvianoFLPPT10,Baral:2003,BrewkaET11,gekakasc12a,Gelfond88thestable,GelKah:2014,Leone:2006,LinZ04,SimonsNS02}
and \fql~\cite{4qloxford,4qljancl,4qlamsta,4qligpl}. 
A detailed comparison of the selected features of~\asp, \fql and \fsp as well as models used in these languages is provided in Tables~\ref{tab:compar} and~\ref{tab:models}. 

\begin{table}
	\caption{A comparison of  \asp, \fql and \fsp features.\label{tab:compar}}
	\centering\begin{tabular}{c||c|c|c|c}\hline
		Language	& Number & Model & Stratification & Consistency 		\\ \hline
		\asp & $\leq$ Exp & NP & \ieeeonly{Not required}\aaaionly{No} & \ieeeonly{Required}\aaaionly{Yes} \\
		\fql & $=1$ & P & \ieeeonly{Required}\aaaionly{Yes} & \ieeeonly{Not required}\aaaionly{No} \\
		\fsp & $\geq 1$, $\leq$ Exp & P & \ieeeonly{Not required}\aaaionly{No} & \ieeeonly{Not required}\aaaionly{No}\\
		\hline 	
	\end{tabular}
\end{table}

\begin{table}
	\caption{A comparison of the discussed models.\label{tab:models}}
	\centering \begin{tabular}{c||c|c|c|c}
		Models & Truth  & Consistency & Minimality & Supportedness\\ \hline
		Stable models  & \true, \false & Yes & Yes & Yes\\
		Answer Sets & \true, \false, \unknown & Yes & Yes & Yes \\
		Well-supported & \true, \false, \unknown, \incons & No & No & Yes \\
	\end{tabular}
\end{table}

Paraconsistent logic programming has been studied by~\cite{BlairS89} who extended Kripke-Kleene semantics investigated in~\cite{Fitting85} to a~four-valued framework founded on Belnap logic $B_4$. The first paraconsistent approach to \asp has been proposed in~\cite{sakama-inoue}, where a~\belnap-based framework is used and extended to six- and nine-valued frameworks for reasoning with inconsistency. Unlike~\cite{sakama-inoue}, we use~\msv together with trial/inspection operators as uniform means for disambiguating inconsistencies and completing missing knowledge in a nonmonotonic manner. For a survey of paraconsistent approaches to logic programming see also~\cite{DP98}.

Paracoherent \asp~\cite{Eiter:2010} aims at reasoning from \asp programs lacking answer sets due to cyclic dependencies of atoms and their default negations. Program~\ref{prog:unk} with Line~\ref{line:unk1} substituted by the equivalent (with respect to \kleene) \asp rule `$p\!\when\!\!\dneg q.$' is an example of such a dependence. For paracoherent reasoning~\cite{Eiter:2010} consider semi-stable models of~\cite{sakama-inoue} and semi-equilibrium models. In \fsp, Program~\ref{prog:unk} has a~single model with both $p$ and $q$ inconsistent, $\{p,\!\lneg p,q,\!\lneg q\}$, the same no matter whether $\dneg q$ is assumed \true or \false. Both, for semi-stable and semi-equilibrium semantics, model generation is proved intractable.

A hierarchy of tractable classes of stable models (over~\kleene) has been reported in~\cite{Ben-Eliyahu96}. It reflects programs distance from their stratifiability. However, one of complexity factors considered there is, in the worst case, exponential with respect to number of propositional variables, what makes it intractable in the framework we consider. 

Sufficient conditions for \asp  guaranteeing tractability of answer set generation have been identified in the literature~\cite{FichteHMW17,FichteS15}. Also, tractable default reasoning subsystems that can be translated into \asp have been considered in many sources, including~\cite{EliyahuD96,EiterL00,Garcia05,Selman:1992}. However, these approaches cover substantial subclasses of the general problem for which we have achieved tractability.


In summary, we have defined the \fsp language combining the \asp and \fql ideas. We have gained  tractability of model generation by relaxing the consistency requirement. To our best knowledge, tractability of model generation for \asp-like languages has not been achieved before. That way, a prevalent use of paraconsistency allowed us to achieve tractability of the, otherwise intractable, problem. 

\fsp is intended to serve as a tool complementary/parallel to \asp, being useful for paraconsistent reasoning and providing models when \asp model generation fails due to  complexity reasons or inconsistencies involved.

\ieeeonly{In an extended version of the paper we we plan to extend \fsp by:
	\begin{itemize}
		\item adding inspection operators to resolve ignorance and disambiguate inconsistencies: when the program will stratifiable tractability is retained, otherwise NP-completeness can be proved;
		\item adding disjunctive conclusions.
	\end{itemize}
}


\end{document}